# A Self-Organized Resource Allocation using Inter-Cell Interference Coordination (ICIC) in Relay-Assisted Cellular Networks

Mahima Mehta[1], Osianoh Glenn Aliu[2], Abhay Karandikar[3], Muhammad Ali Imran[4]

[1,3]*Department of Electrical Engineering,*
*Indian Institute of Technology Bombay, Mumbai, India*
*E-mail: mahima, karandi @ ee.iitb.ac.in*
[2,4]*CCSR, University of Surrey, Guildford, UK*
*E-mail: o.aliu, m.imran @ surrey.ac.uk*

**ABSTRACT:**

*In a multi-cell scenario, the inter-cell interference (ICI) is detrimental in achieving the intended system performance, in particular for the edge users. There is paucity of work available in literature on ICI coordination (ICIC) for relay-assisted cellular networks (RACN). In this paper, we do a survey on the ICIC schemes in cellular networks and RACN. We then propose a self-organized resource allocation plan for RACN to improve the edge user's performance by ICIC. We compare the performance of reuse-1, reuse-3, soft frequency reuse (SFR) scheme, proposed plan with and without relays. The performance metrics for comparison are edge user's spectral efficiency, their signal-to-interference-and-noise ratio (SINR) and system's area spectral efficiency. We show by the simulation results that our proposed plan performs better than the existing resource allocation schemes in static allocation scenario. Next, we propose to make our resource allocation plan dynamic and self-organized. The distinct features of our proposed plan are: One, it achieves a trade-off between the system's area spectral efficiency and the cell edge spectral efficiency performance. Secondly, it introduces a novel concept of interfering neighbor set to achieve ICIC by local interaction between the entities.*

## 1. INTRODUCTION:

In conventional cellular systems, static resource planning approach was followed in which a fixed set of resource was allocated to cells. However, with increasing temporal and spatial variations of traffic, situations often arise when few cells happen to starve for spectrum while in others, spectrum remains unused. As a consequence, set of users in the former case will have higher call blocking probability due to paucity of resources. In the later case, there is inefficient resource utilization due to plethora of resources remaining underutilized. Thus, in a variable traffic scenario, static resource planning will be inefficient. Hence, to alleviate this unbalanced resource distribution, a flexible resource planning is required which dynamically varies resource allocation as per the traffic. A classical paper [1] gives a comprehensive survey on the evolution of various resource planning schemes based on the changing scenarios from conventional to the present times. It emphasizes the impact of increase in traffic, demand for high-bandwidth applications and interference on resource planning.

The resource planning domain is benefitted by adapting orthogonal frequency division multiple access (OFDMA) as multiple access mechanism (recommended by third generation partnership project – Long Term Evolution (3GPP-LTE standard) [3], [4]. The resource allocation in OFDMA ensures that no two users are assigned a common resource in a cell at a given time [2], thereby eliminating intra-cell interference (due to transmissions within the cell). Now, main research focus is on inter-cell interference (ICI). ICI is due to transmissions from outside the cell. It is detrimental in achieving the intended system performance, particularly for the users located close to cell boundary, henceforth referred to as edge users. One of the approaches being considered in 3GPP-LTE to resolve this problem is interference avoidance/ coordination (ICIC) [5]. Its objective is to apply restrictions to the resource allocation by coordination between network entities [6]-[12] so that ICI is minimized. Thus, resource allocation plans with ICIC offers performance improvement for edge users in an OFDMA-based cellular network.

Relaying is one approach to improve edge user's performance. In addition, it facilitates ubiquitous coverage and better capacity [13]-[14]. The wireless fading channel due to its multipath nature can cause the received signal quality of users to fall below the acceptable limits. Such users are then said to be in outage [15]-[16]. A user can be in outage irrespective of its location (close or far off from transmitting node). Relay deployment benefits both users on edge and in outage. However, it adds one more dimension of complexity in resource planning [17], [18] due to the need of resource sharing and information exchange between relay node (RN) and base station (known as Evolved NodeB/ eNB as per 3GPP standards).

Thus, relaying makes ICI mitigation more challenging [19]. In this paper, we address this problem of ICI in an OFDMA-based relay-assisted cellular network (RACN).

Relays can also play a significant role in making the system self-organized. Consider a scenario when system can sense the environment autonomously and then, resource allocation algorithm adapts to the variations that were sensed. This leads to self-organization which is envisaged to play a key role in the next generation cellular networks [20]. It relies on local interaction between entities (eNBs and RNs) in order to adapt the algorithm to meet the intended performance objectives. The resource planning for cellular systems thus becomes more involved.

With an objective of ICI mitigation in OFDMA-based cellular networks, various policies have been proposed in the literature as – static frequency reuse schemes [24]-[25] like fractional frequency reuse (FFR), power control based reuse schemes like soft frequency reuse (SFR) [21]-[22], the variants of SFR as SerFR [23] and modified SFR (MSFR) and dynamic resource plans [26]-[30]. Researchers have also used different approaches for resource planning and interference management as reinforcement learning, Q-learning [31]-[33], cognitive radio [32] and self-organization [34-35]. The resource planning for RACN is discussed in [17], [37]. However, the literature has limited contributions in ICI mitigation in RACN [38]-[40] which mostly rely on different reuse schemes to alleviate ICI.

In the light of contributions so far, we are motivated to address the challenges imposed by relaying. To the best of our knowledge, self-organized resource plans have not been implemented in RACN scenario. In this paper, we present a framework for a self-organized resource allocation plan with ICIC for the OFDMA-based RACN. The expected outcomes of our proposed solution are: efficient resource utilization, improved edge user's performance and flexibility and adaptability to optimize the resource allocation algorithm according to the variations in environment. In our solution, we facilitate flexible resource sharing between eNBs and RNs such that any resource can be used in any region unless interference exceeds the acceptable threshold. Based on this localized rule, resources will be dynamically shared between the set of interfering neighbors such that no two adjacent cells use same co-channels. This will achieve ICIC in RACN. This is an extension of the initial work done in [20] to demonstrate the self-organized, distributed and dynamic resource allocation in a cellular network.

The rest of the paper is organized as follows. In Section II we give an overview of the OFDMA-based cellular networks, discuss the impact of ICI and the recommendations given by 3GPP-LTE standard. Then, various resource allocation schemes proposed in the literature to mitigate ICI are reviewed in Section III as static and dynamic resource allocation plans and self-organized resource allocation schemes. Finally the scenario in RACN is reviewed. In Section IV, we describe the system model and the algorithm of our proposed self-organized resource allocation plan for an OFDMA-based RACN. The simulation results are discussed in Section V. In Section VI, we give the conclusions and future work.

## 2. Overview of an OFDMA-based Cellular Network and the problem of Inter-Cell Interference (ICI):

The ability of Orthogonal Frequency Division Multiplexing (OFDM) to combat frequency-selective fading makes it a suitable candidate for modulation in the next generation wireless communication. OFDM transforms the wide-band frequency-selective channel into several narrow-band sub-channels and transmits the digital symbols over these sub-channels simultaneously. Then, each sub-channel appears as a flat fading channel. This makes the system robust to multipath fading and narrowband interference [16].

In a multi-user environment, each sub-carrier will exhibit different fading characteristics to different users at different time instants. It will be due to the time-variant wireless channel and the variation in users' location. This feature can be used to our advantage by assigning sub-carriers to those users who can use them in the best possible way at that particular time instant. Such an OFDM-based multiple-access scheme is known as OFDMA. It allocates a set of sub-channels[†] or sub-carriers to users exclusively for a given time instant. The minimum set of sub-carriers that are assigned for a certain fixed time-slots is known as a resource block (RB) or chunk. The composition of RB is a design issue. In addition to the sub-carrier allocation, other resources as power and modulation scheme can also be assigned on per sub-carrier basis to each user. Thus, OFDMA facilitates a flexible resource planning due to the granularity of the resources available for allocation, for example, low and high rate users can be assigned a small and a large set of sub-carriers respectively with certain power and modulation settings. With the increasing number of users, more will be the choice of users who can best utilize a given sub-carrier. This is known as *multi-user diversity* [15]-[16]. To exploit this feature of OFDMA, it is required to have a resource allocation scheme which adapts to the changing channel conditions experienced by users on temporal basis. It is known as an *adaptive* resource allocation scheme.

---

[†] A sub-channel may be defined as a set of sub-carriers. However, we will not differentiate between the two terms in this paper.

From the perspective of radio resource management, the performance of OFDMA-based cellular system can have following three optimization policies [4]:

- *Subcarrier selection for users*: It determines the set of subcarriers with high signal to noise and interference ratio (SINR) for assignment to the users in a time slot. This ensures high data rate transmission and maximizes the system's instantaneous throughput.
- *Bit loading*: In downlink (DL), eNB determines the modulation and coding scheme (lower or higher level) to be used on each sub-carrier. This decision is based on Channel Quality Indicator (CQI), which is an indicative of data rate that can be supported by DL channel (determined by SINR and receiver characteristics).
- *Power loading*: It determines the amount of power on each subcarrier. This helps offer variable power allocation to different group of subcarriers to optimize its usage.

All the above mentioned optimization policies depend on channel condition and therefore channel estimation needs to be accurate. The adaptive resource allocation can have any combination of the above three optimization policies.

Based on the objective function, the approaches for resource allocation schemes in OFDMA can be categorized into two types: one, *System-centric approach*, where the objective is to optimize the metrics as data rate and transmission power. This approach does not consider user's achievable performance and may lead to unfairness. For example, opportunistic scheduling maximizes system throughput at the cost of being unfair to the users with poorer channel condition [16]. The other is *Application-centric approach* which sets the objective from user's perspective and aims at maximizing *utilities* like fairness, delay constraints etc. Each user can have its own utility function for a certain resource and the objective is to do resource allocation to maximize the average utility of system. An overview of different allocation schemes is given in [2] with different objectives as maximizing throughput, minimizing power consumption or optimizing certain utility function etc.

In a multi-cell environment, *edge users* experience the greatest amount of degradation in system performance due to *inter-cell interference* (ICI). The transmit power falls off with distance and therefore received signal strength at the cell edge is low. Being located closer to the cell boundary, edge users are prone to interference from eNB's in the neighboring cells that use the same RBs in DL. As a consequence, they experience low SINR and therefore require more RBs and higher transmit power compared to other users to meet the same data rate requirement. This consumes more resource and reduces system throughput as well. Thus, edge users are served at a cost of resource utilization efficiency and system throughput. This trade-off between the maximization of system's throughput and spectral efficiency and improving the edge user's performance is addressed by using a variety of frequency reuse plans [23]-[24], [28]-[29]. Yet another approach to mitigate ICI is to observe the system as collision model where ICI is treated as collision [25]. The objective is to reduce collision probability and improve capacity by either restricting the usage of RBs in cells or by reducing the transmit power of the RBs lying in collision domain. Efficient resource planning is therefore essential to mitigate ICI, improve edge users' throughput and simultaneously improve resource utilization. The next sub-section briefly mentions the recommended schemes for handling ICI in 3GPP-LTE standard, followed by a discussion on the issues of concern in interference coordination schemes.

## 2.1. Recommendations for mitigating ICI in 3GPP-LTE

Following approaches are recommended by 3GPP-LTE standard [3] for interference mitigation in OFDMA-based cellular networks:

- *Interference randomization*: It includes cell-specific scrambling, interleaving, and frequency hopping.
- *Interference cancellation*: It can be done in two ways, one is to detect interference signals and subtract them from received signal. The other involves selecting the best quality signal by suitable processing. This is applicable when multiple antennas exist in system.
- *Interference avoidance/coordination*: This scheme controls the resource allocation by coordination between network entities [6]. Details follow in next Section.
- *Adaptive beamforming*: It is used for ICI mitigation in DL, where antenna can adaptively change its radiation pattern based on the interference levels. Though it complicates antenna configuration and network layout, but the results are effective.

The methods of interference avoidance/coordination and adaptive beam forming are very promising from the perspective of improving edge user's performance. Therefore, both are being preferred for deployment in the 3GPP-LTE systems. We illustrate coordination-based scheme for ICI mitigation in next sub-section.

## 2.2. Inter-Cell Interference Coordination (ICIC)

The basic concept of ICIC is to restrict the usage of resources (time/frequency and/or transmit power) such that the SINR experienced by edge users increases and their achievable throughput improves. First, it determines the resources available i.e. the bandwidth and power resources in each cell. Then, it determines the strategy to assign them to users such that ICI remains below the acceptable limits. ICIC has been widely investigated for LTE systems [7].

The issues of concern in inter-cell interference coordination (ICIC) are:

- The information exchange between network entities will ensure coordination in resource allocation decision. However, the amount of overheads involved will require extra processing and will either consume the scarce frequency resource or will require backhaul link for communication [41]. For example, LTE provisions to modify power settings based on the performance indicators in DL and interference indicators in uplink (UL) which are exchanged over the X2 interface (signaling interface between eNBs in LTE). The performance indicator for DL can be Relative Narrowband Transmit Power (RNTP) per PRB and the interference indicators in UL are High Interference Indicator (HII) and Overload Indicator (OI) as specified in the LTE standards [42]-[43].
- To ensure interference avoidance, sub-channels with high amount of interference will not be used for allocation, even if their channel state is good [5]. This will lead to under-utilization as well as inefficient utilization of resources. Also, multi-user diversity (i.e. assigning sub-channels only to users who can achieve the maximum possible channel capacity) cannot be exploited well in such a case even though the channel is frequency-selective.
- As the channel condition is time-varying, parameters of resource management algorithm needs to be updated periodically, which requires more resources for feedback and signaling.
- This coordination-based strategy will essentially maximize system throughput by minimizing ICI, but it may lead to some amount of unfairness to the users [5]. Thus, fairness in allocation is also to be considered.

To summarize, the basic motive behind any ICIC mechanism is to either avoid allocating those RBs that are interfering or to use them with lower power levels [15]. Different resource allocation schemes with ICIC proposed in the literature are reviewed in next Section.

## 3. Overview of Resource Allocation Schemes in OFDMA-based Cellular Networks:

The resource allocation schemes can be broadly classified into two categories: static and dynamic. The static allocation schemes utilize the fact that edge users need a higher reuse as they are more prone to ICI compared to cell-centre users. These schemes rely on fractional reuse concept, i.e. users are classified based on their SINR which is an indicative of ICI they experience. Then, different reuse patterns are applied to them based on their experienced level of interference. However, resources allocated for cell-centre and edge users are fixed. The static ICIC schemes have lower complexity and lesser overheads. Next sub-section illustrates these schemes.

### 3.1. Static Resource Planning

An interesting fact that governs cellular system design is that the signal power falls diminishes with distance. This feature helps in ensuring efficient resource utilization. It allows frequency resource to be reused at a spatially separated location such that signal power diminishes to the extent that it does not cause any significant interference. The distance at which the frequency resource can be reused is known as *reuse distance* and this concept is known as *frequency reuse*. The interference due to this reuse is known as *inter-cell* (also known as *co-channel*) *interference.*

In *universal frequency reuse* or reuse-1 (Figure 1a), ICI is high because the reuse distance is 1. The frequency resource is utilized well as all RBs are available in each cell, albeit the edge users are prone to more interference because the RBs are reused by adjacent cells. To reduce this interference, the reuse distance is to be increased. With frequency reuse concept, each cell will now have only a fraction of the resource and hence available RBs in a cell will reduce. As an example, reuse-3 is shown in Figure 1b. However, this reduction in resource availability is compensated by the fact that edge users will not get interference from adjacent cells which will improve their throughput.

The significant point to note here is that the edge users are more prone to ICI compared to the cell centre users and therefore if higher reuse is deployed only for the edge users, we can achieve a trade-off between resource utilization and ICI mitigation. Thus, in mitigating ICI, frequency reuse scheme can be made fractional to ensure that a certain part of the allocated spectrum is reserved for edge users. This improves data rate and coverage for cell edge users [8] and also ensures fairness. The channel

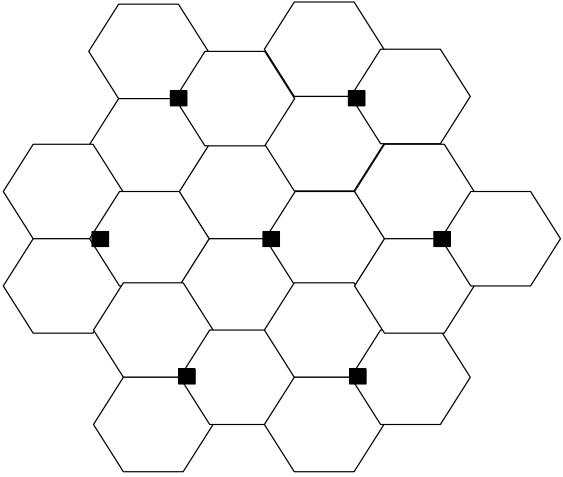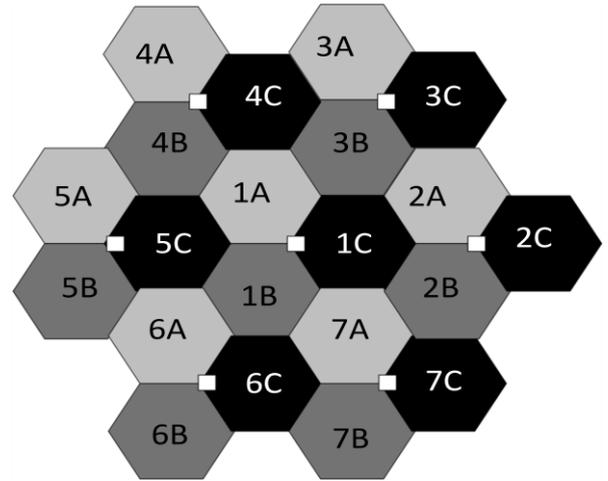

| Figure 1: a. Frequency Reuse-1 | Figure 1: b. Frequency Reuse-3 |

*partitioning schemes* are introduced to achieve this trade-off and improve the system performance. Higher reuse factor eliminates co-channel interference from adjacent cells and improves the SINR. It has been shown in [21] that for reuse-3, the gain in SINR compensates for the loss in bandwidth due to fewer channels available in cell thereby improving the overall channel capacity. However, for reuse more than 3, this compensation does not take place and hence channel capacity reduces.

In a *Fractional Frequency Reuse* (FFR) scheme, available RBs are partitioned into two sets: inner set to serve cell-centre users[†] (closer to eNB) and outer set to serve edge users. It primarily allocates resources with a higher frequency reuse to edge users and with reuse-1 to the cell-center users so that effective reuse is greater than 1. For example, in *Partial Frequency Reuse* (PFR) [25], total available RBs are partitioned into two sets, one for cell-centre users (with $C$ resource blocks) and other for edge users (with $E$ resource blocks), where central-band has reuse-1 and the edge band has reuse-3. The number of resource blocks/cell in this case will be $C + E/3$.

Many variants of reuse schemes have also been proposed in the literature. In [24], authors show that with a-priori FFR planning, spectral efficiency can be improved. Researchers have demonstrated that ICIC is achieved using FFR which helps in improving performance of edge users [27] as well as maximizing throughput [26].

In a nutshell, these schemes are based on allocating a certain fixed number of RBs in a cell, which essentially hard limits the achievable user throughput because only a portion of bandwidth is made available in the cell.

[†]: Discriminating users as cell-centre or cell-edge can be a function of distance, SINR or achievable throughput etc.

This issue becomes significant when there is spatially-distributed heterogeneous traffic load. Thus, in spite of various FFR schemes proposed in the literature, the recurring challenge is *limiting throughput and low spectral efficiency*. To resolve these problems, FFR/PFR can be made more efficient by dynamically changing the reuse factor so that capacity and performance improves compared to static FFR schemes. Such dynamic reuse schemes are discussed in next sub-section.

## 3.2. Dynamic Resource Planning

One such scheme which does power control along with dynamically changing the reuse factor is *Soft Frequency Reuse* (SFR) [21]-[22]. In SFR, total RBs are divided into three set of sub-bands and all are made available in each cell (Figure 2) such that cell centre users have reuse-1 while cell edge users have reuse-3 or more [9]-[12]. This is known as *soft reuse* because the channel partitioning applies only to edge users while cell-centre users have the flexibility of using the complete set of RBs, but with lower priority than the edge users. There is one maximum permissible transmit power level set for both cell-centre users and edge users such that the maximum permissible transmit power for edge users is higher than the one for cell-centre users. The ratio of transmit power of edge users to that of cell-centre users is known as *power ratio* and adjusting this ratio from 0 to 1 will vary the effective reuse from 3 to 1 [21]. Thus, SFR is a trade-off between reuse-1 and reuse-3. This power ratio can be adapted based on the traffic distribution in a cell, for example, power ratio will be low when user density on cell-edge is high, and will be higher when user density is high in cell-centre.

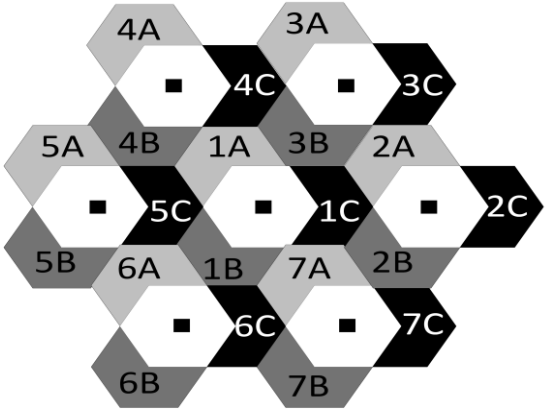

Figure 2: Soft Frequency Reuse (SFR)

Thus, SFR [21]-[22] allows each cell to utilize full bandwidth and thus maximize resource utilization efficiency. In [28], capacity comparison for SFR and PFR with reuse-1 is done and it is shown that SFR enhances cell-edge throughout without sacrificing average cell throughput. To achieve this, it needs to do a perfect power control on RBs and mitigate ICI. Its implementation requires careful coordination between the entities by exchanging relevant information (overload, interference indicators etc.) and adjusting the number of RBs and their power allocated in a cell so that ICI can be mitigated by coordination. To summarize, efficient implementation of SFR requires coordination between adjacent cells and cooperative resource allocation without any central controlling entity. This is the way a self-organizing network (SON) is envisaged to operate. Mitigating ICI by coordination (ICIC) thus fits within the framework of self-organized cellular networks.

In [23] the downsides of SFR are highlighted as large frequency-selective scheduling gain loss and low peak rates for edge users. This is due to the fact that edge users get only a fraction of resources available. Then, selection of best resource-user combination for allocation is done from only a subset of RBs while there could be other RBs offering better achievable throughput which are not available in the subset. Also, it is shown that it is difficult to ensure maximum sector throughput and edge user throughout simultaneously. To address this issue, authors proposed a softer reuse (SerFR) scheme in which reuse factor for both cell-centre and edge users is 1 and a modified proportional fair scheduler is used which gives preference to edge users over cell-centre users and also ensures fairness amongst them. It is thus essential for resource management algorithms to adapt to system dynamics while keeping the flexibility of using entire spectrum resource in every region. The insight is to keep the resource planning adaptive with no inherent constraints from design perspective. A modified SFR (MSFR) scheme is proposed in [36], which introduces SFR into the "pre-configured and Fixed (PreF)" allocation scheme and shows significant performance improvement.

In general, dynamic reuse plans tend to perform better than their static counterparts due to the fact that they provide the flexibility of using the complete resource set. The dynamic resource plans for interference mitigation are proposed in [29], [32]. In [31], authors use reinforcement learning for dynamic resource planning. The generation of soft-FFR patterns in self-organized manner is focused in [34]-[35] where resource allocation (i.e. determining number of sub-carriers and power assignment) is performed by dynamically adapting to the traffic dynamics for constant bit rate (CBR) and best-effort traffic. They have compared the performance for two cases - without and with eNB's coordination and showed that performance is better with coordination. In next Section, we review the resource planning and ICI mitigation schemes in RACN.

### 3.3. Resource Planning in RACN

Users (also known as user equipments (UEs) as per the 3GPP-LTE standard) in outage or on edge are benefited when relay nodes (RNs) assist eNBs in their transmission due to two reasons: one, RN has higher receiver antenna gain which makes low power transmission by eNB feasible and secondly, RN can also transmit with low power due to its proximity to UE. Thus, relay deployment brings down power consumption in DL, reduces interference and ameliorates system performance [13].

One of the major challenges in relay deployment is that of resource sharing between eNB and RN. Two basic frequency plans exist for such networks: one, in which eNB and RN have disjoint spectrum allocation (*orthogonal allocation*) and other, in which spectrum is shared between the two (*co-channel allocation*) [13]. The former reduces interference due to orthogonal allocation but available resource with each node also reduces by the same amount which makes resource utilization inefficient. Therefore, later case of sharing frequency is a more viable option as more resources are available and by proper interference management, system's performance can be improved.

However, there is limited literature available which addresses the problem of interference management in RACN, compared to that in single-hop OFDMA-based cellular networks (discussed in sub-section 3.1 and 3.2). An overview of radio resource management issues in RACN is given in [17]. In [37], authors propose a dynamic frequency reuse scheme for wireless relay networks where orthogonal frequency allocation is done to relays (which are randomly located) within the cell. A dynamic score based scheduling scheme is proposed in [38] which considers both throughput and fairness and achieves performance improvement in terms of SINR and edge user's throughput. It uses combination of static and dynamic allocation. In [39], authors have divided the frequency resource into two zones: inner and outer correspondingly for eNB and RNs. They use directional antennas and specific frequency bands to eliminate ICI. Their scheme is shown to perform better that MSFR proposed in [36] in terms of average spectral efficiency. Next Section discusses our proposed self-organized resource allocation scheme with ICIC in RACN which has not been addressed so far in the literature.

## 4. System Model

Consider a two-hop fixed RACN with OFDMA as multiple access technique. For cellular deployment, we use a clover-leaf system model (Figure 3a) where each cell site comprises three hexagonal sectors with one eNB per cell located at the common vertex of these three sectors. The hexagonal geometry of sectors makes mathematical analysis simpler. The motivation for clover-leaf model is that it appropriately demarcates the radiation pattern of a cell site utilizing three sector antennas. There is one RN in each sector (Figure 3b) placed on cell edge. Both eNB and RN deploy a tri-sector antenna. As shown in Figure 3b, the three RN antennas will be serving users located in regions 1A, 1B and 1C respectively.

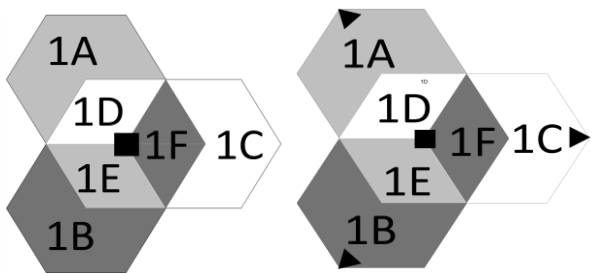

Figure 3: Single cell of clover-leaf model with eNB at the centre: Proposed System Model (a) without Relays (b) with Relays on the cell edge

"Multihop" is a generalized term for RACN that implies presence of more than one relay node between eNB and user. It involves issues like route selection in addition to resource allocation. However, to investigate performance improvement in a multi-hop cellular system, it is a reasonable assumption to consider two-hop scenario, i.e. only one RN between eNB and user. As verified in [23] maximum throughput gains for multihop networks is obtained with two or three hops. Hence, we consider a two-hop OFDMA-based cellular system to implement the proposed algorithm for DL transmission scenario.

A few terminologies introduced in our algorithm are mentioned below:

*Classifying Regions*: We call the region of cell-centre users as *non-critical region* (indicatively inner hexagon, i.e. regions labeled 1D, 1E and 1F in Figure 3a). We give this name because users in this region are less prone to ICI. Correspondingly, we call the region of edge users as *critical region* (indicatively, regions labeled 1A, 1B and 1C in Figure 3a) as users in this region are vulnerable to ICI. In our system model, we deploy reuse-3 for both categories of users and therefore there is a critical and a non-critical region in each sector (Figure 3a).

*User classification:* Users are uniformly distributed in each sector with random locations. Based on signal-to-noise ratio (SNR), we classify them as *Non-Critical users* (cell-centre) and *Critical users* (edge users). This decision is based on threshold value of SNR e.g., users with estimated SNR less than 25th percentile of the whole system are regarded as critical users and others as non-critical users. This threshold is a design parameter. Non-Critical users are close to serving eNB experiencing high SINR and therefore demanding fewer resources. Critical users are those who experience low SINR and therefore demand more resources. They are also one of the dominant sources of interference (as being away from eNB, their transmission requires large amount of power).

*Association Identification:* To determine serving node for a user, we follow a rule that all non-critical users are served by eNB and all critical users by RNs of their respective sector.

*Interfering Neighbor set*: This is motivated by the concept of *sectorial neighbors* discussed in [20] for a simple cellular system model without relays. The *sectorial neighbors* are the set of adjacent sectors from neighboring cells sites (Figure 4) which are considered to cause interference. The adjacent sector of the same cell is not considered because it is assumed that there is no intra-cell interference.

We extend this concept of *sectorial neighbors* to a scenario when RNs are deployed in system. It will involve identifying interferers for users in every region. It is because with RNs in system, each sector has a critical and non-critical region and users in every region will encounter interference from a different set of transmitting nodes. The *interfering neighbor set* comprises that set of adjacent regions, which will cause interference (when transmission is done to users in these regions) based on directivity of antennas at eNB/RN and co-channel usage.

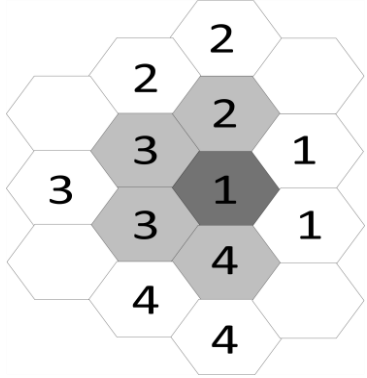

Figure 4: Sectorial neighbor concept [20]

The interfering neighbor sets will be indicated in the Neighbor Matrix N given by –

$$N = \{n_{i,j} | n_{i,j} \in {0,1}\}_{R \times R} \quad , \quad (1)$$

where

$$n_{i,j} = \begin{cases} 1, \text{ region } i \text{ interferes with region } j \text{ and } i \neq j \\ 0, \text{ region } i \text{ does not interfere with region } j \\ \quad \text{ and cochannels can be allocated.} \end{cases}$$

This neighbor matrix will be used as a look-up table to determine the set of interfering nodes in every transmission time interval (TTI).

To justify the impact of our proposed scheme in interference mitigation, we compare performance of our proposed resource allocation scheme (for two cases: without and with relays) with the existing schemes of reuse-1, reuse-3 and soft frequency reuse (SFR). The performance metrics used for comparison are SINR, spectral efficiency of edge users and system's area spectral efficiency. They are illustrated in following sub-sections.

### 4.1. SINR Measurement

Our reference cell is centre cell for which interference will be considered from the first tier of cells. Note that our algorithm is for DL resource allocation case. Therefore, interference will be from eNBs and/or RNs only.

To evaluate path loss, macro cell propagation model of urban area is used as specified in [45], where $L$ is path loss and $R$ is distance (in Km) between eNB and user.

$$L = 128.1 + 37.6 \, log_{10} R. \quad (2)$$

In conventional universal frequency reuse, every other node $c$ transmitting in same transmission time interval (TTI) would serve as interference. The corresponding SINR of each user will be -

$$SINR_{FR1}(u) = \frac{P_u \times \xi_u}{N_o \Delta f + \sum_{c \neq u} P_c \times \xi_c} \quad (3)$$

where $u$ is a user in reference cell. $P$ is transmit power, $\xi$ is log-normal shadowing with mean 0 and standard deviation $\sigma_{eNB}$ for eNB-UE link, $N_0$ is noise spectral density and $\Delta f$ is user bandwidth.

However for FFR scheme, each sector of cell is given a fixed portion of total RBs and same pattern is followed all through the network. This reduces interference experienced from other cells as adjacent sectors of other cells do not interfere with each other. Using reuse-3, the SINR is calculated as -

$$SINR_{FR3}(u) = \frac{P_u \times \xi_u}{N_o \Delta f + \sum_{c \neq u, c \in F} P_c \times \xi_c} \quad , \quad (4)$$

where $F$ is a set of RBs used by user $u$.

In SFR scheme [7], transmission is done to critical users with higher power and to non-critical users with lower power. RB allocation is done to the critical users on higher priority with reuse-3 and non-critical users are free to use any RB but with lower priority than the critical users. This scheme facilitates using any RB anywhere but with predetermined priorities and appropriate power levels.

Let the ratio of number of edge users to cell-centre users be $\alpha_U$ and the ratio of transmit power for edge users to that of cell-centre users (power ratio- described in sub-section 3.2) be $\alpha_P$. Now, transmit power ratio $\alpha_P$ will be adaptively varied based on user density ratio $\alpha_U$.

The *SINR* for cell-centre user is expressed as -

$$SINR_{SFR}(u\_cc) = \frac{P_{u\_cc}^{cc} \times \xi_{u\_cc}}{N_o \Delta f + \sum_{c \neq u\_cc, c \in F} P_c^{cc} \times \xi_c} \quad (5)$$

The *SINR* for edge user is expressed as -

$$SINR_{SFR}(u\_eu) = \frac{P_{u\_eu}^{eu} \times \xi_{u\_eu}}{N_o \Delta f + \sum_{c \neq u\_eu, c \in F} P_c^{eu} \times \xi_c} \quad , \quad (6)$$

where $u\_cc$ is cell-center user, $u\_eu$ is edge user, $P^{cc}$ is transmit power for cell-center users and $P^{eu}$ is transmit power for edge user. The transmit power levels ($P^{cc}$ and $P^{eu}$) must satisfy the power ratio $\alpha_P$, which is given by $\alpha_P = \frac{P^{eu}}{P^{cc}}$ and power ratio itself is determined according to user density ratio $\alpha_U$ as mentioned below –

$$\alpha_P = \begin{cases} \frac{1}{3} & \text{if } \alpha_U \geq 50\% \\ 1 & \text{if } 50\% > \alpha_U \geq 25 \\ \frac{3}{1} & \text{if } \alpha_U < 50\% \end{cases} \quad , \quad (7)$$

where $\alpha_U = \frac{N_{EU}}{N_{CC}}$, $N_{EU}$ is number of cell edge users and $N_{CC}$ is number of cell-centre users.

This '*user density based transmit power adaptation*' in SFR helps in improving edge user's performance.

*Interference Analysis in proposed scheme without relays:*

In our proposed scheme without relays, the set of RB allocation is done such that disjoint set of RBs are assigned to edge and cell-centre users in every sector. Based on SNR threshold, a user is identified as an edge or a cell-centre user. Unlike SFR, there is no '*user density based transmit power adaptation*'. Instead, we use two fixed power levels, $P_{high}$ for edge users and $P_{low}$ for cell-centre users.

SINR for a user will be computed as -

$$SINR_{PRA\_ORN}(u) = \frac{P_u^T \times \xi_u}{N_o \Delta f + \sum_{c \neq u, c \in F} P_c^T \times \xi_c} \quad (8)$$

where $P^T = \begin{cases} P_{high}, & \text{if estimated SNR < Threshold} \\ P_{low}, & \text{if estimated SNR > Threshold} \end{cases}$

and $SINR_{PRA\_ORN}(u)$ is SINR of user $u$ in the proposed resource allocation scheme without RNs in the system.

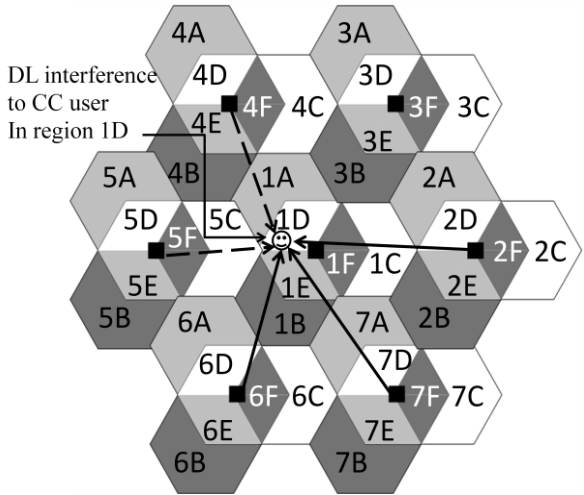

Figure 5: Interference scenario in the proposed scheme (without relays) for cell-centre user

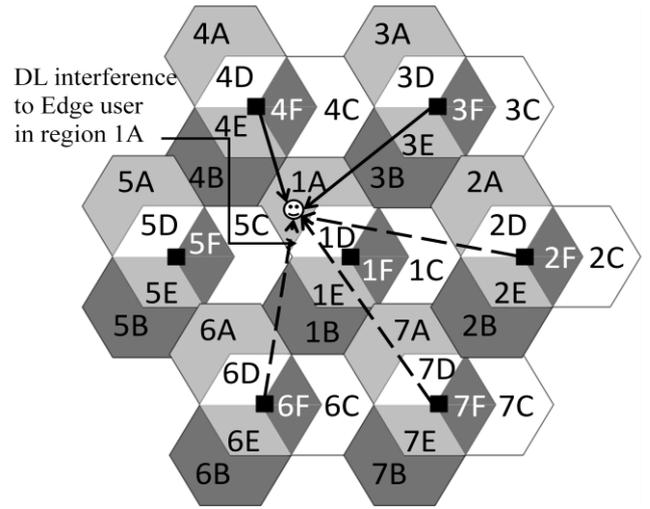

Figure 6: Interference scenario in the proposed scheme (without relays) for cell-edge user

The set of interfering nodes will be different for both user categories as shown in Figure 5 and 6. For example, a cell-centre user (indicatively located in region 1D) will face interference from eNBs 2, 6 and 7 with their transmit power level set to $P^{cc}$ and also from eNBs 4 and 5 with their transmit power level set to $P^{eu}$ (Figure 5). Similarly, for an edge user (indicatively located in the region 1A), interference will be from eNBs 2, 6 and 7 with their transmit power level set to $P^{eu}$ and also from eNBs 3 and 4 with their transmit power level set to $P^{cc}$ (Figure 6). This can also be extended for any network size.

*Interference Analysis in the proposed scheme with relays:*

In this scenario (with relays in our system model), we will be able to address the problem of capacity, coverage and further improvement in edge user's performance jointly (Section 4.1). Now, the identified edge users will be served in two hops via RN. Instead of power adaptation, there will be a fixed transmit power for both eNB and RN as specified in the simulation parameters given in Table 1.

$$SINR_{PRA\_RN}(u) = \frac{P_u^T \times \xi_u}{N_o \Delta f + \sum_{c \neq u, c \in F} P_c^T \times \xi_c} \quad (9)$$

where $P^T = \begin{cases} P^{eNB}, & \text{for cell-centre users} \\ P^{RN}, & \text{for cell-edge users} \end{cases}$

and $SINR_{PRA\_RN}(u)$ is SINR of user u in the proposed resource allocation scheme with RNs in the system.

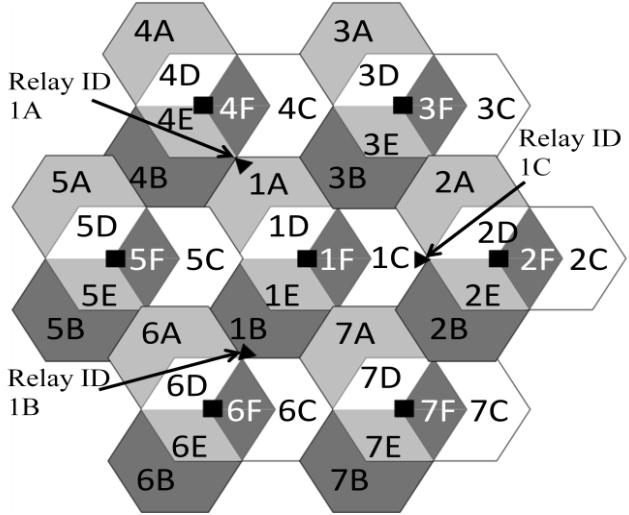

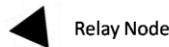

Figure 7: Interference scenario in the proposed scheme (with relays) for cell-centre and the edge users

The interference scenario for cell-centre and edge users is described in Figure 7. The set of interfering nodes change in this case due to additional directional relay antennas deployed. For example, let's consider an edge user located in region 1A. On DL, this user would get interference from only eNBs 3 and 4 and also from RNs 3A, 4A and 5A. Similarly a cell-centre user in region 1D will get interference from only eNBs 2, 6 and 7 and from RN 1C.

### 4.2. Spectral Efficiency of Edge Users

Spectral efficiency is one of the significant metrics to be considered in design of wireless communication networks. Spectral efficiency is measured as the maximum achievable throughput (bits per sec.) per unit of bandwidth. Its unit is bits/sec/Hz. For all the spectrum reuse schemes discussed above, we have computed spectral efficiencies for edge users as

$$\eta = \sum_{u \in E} \log_2(1 + SINR_u) \quad (10)$$

where **E** is the set of edge users in system. The comparative plots are shown in Figure 13.

### 4.3. Area Spectral Efficiency of the system

Asides the spectral efficiency, another key metric to operators in classifying the performance of their network is area spectral efficiency. It focuses on spectral efficiency achieved in a given area. The area spectral efficiency is the measured throughput per hertz per unit area for a given cell resource [15]. This gives a practical representation of the improvement in capacity achieved relative to cell size (and reuse distance) with available resources. If reuse distance is increased, available resource per unit area becomes lesser and hence, resource utilization efficiency reduces. However, it reduces ICI and improves system throughput. Thus, we understand area spectral efficiency as a metric that trades-off efficient resource utilization and throughput maximization (by ICI reduction).

This is one of the significant performance metric [44] to compare different frequency planning schemes which certainly impacts cellular system design. This determines achievable system throughput per unit of frequency per unit area. (bits/sec/Hz/m$^2$). It is computed as-

$$\eta_A = \sum_{r \in R} \frac{\sum_{u \in A} \Delta f \log_2(1 + SINR_u)}{W_r A_r} \quad (11)$$

where **A** is set of all users in the system, **R** is set of all regions, $W_r$ is total bandwidth in region r and $A_r$ is area of any region r. The comparative plots of area spectral efficiency are given in Figure 14.

## 5. Proposed Scheme: Self-Organized Resource Allocation using modified FFR with ICIC

We propose a resource allocation scheme for DL transmissions in an OFDMA-based RACN. Its objective is two-fold: first, to do resource allocation with the motive of minimizing ICI by coordination. The second objective is to make the resource allocation algorithm self-organized by making its allocation autonomous and adaptive, involving interaction with the environment. Our solution is expected to improve cell edge users' performance as well as system's area spectral efficiency.

This scheme relies on two concepts: One is the fact that edge users and cell-center users are to be treated distinctly in mitigating interference due to the former being more vulnerable to ICI. Second concept is to avoid proximity of co-channel reuse by local coordination and by applying restrictions in reusing the resources.

We deploy a modified fractional frequency reuse (FFR) in our algorithm. The distinct feature of FFR is that it has a higher reuse for edge users compared to cell-centre users, so that the edge users in neighboring cells operate on orthogonal channels and there is minimum ICI. However, FFR addresses this problem of ICI at the cost of offering fewer resources in cell-edge region. The proposed scheme in [36] partitions the resources available for edge users while keeping reuse-1 for cell-centre users. The scheme in [39] does resource partitioning for both cell edge and cell-centre users with reuse-6 and reuse-3 respectively. In our paper, we deploy a modified FFR scheme (Figure 3a) for resource partitioning for both user categories such that every region gets one-third of resources, unlike [39] where each partition in critical region gets only one-sixth of the resources. In our proposed scheme, resources are shared to serve both cell-centre and the edge users such that the flexibility of using any resource anywhere remains. The only constraint in this flexible resource sharing is that interference due to usage of any RB must be below the acceptable threshold. We compensate for the reduction in amount of resources available (which reduces by a factor of 1/3) by improving edge user's performance. It is justified to deploy reuse-3 because it is optimal for cell-edge and gives better channel capacity compared to reuse-1 and beyond reuse-3 channel capacity begins to decrease as verified in [21]. Also, we use only three relays per cell to provide for coverage and capacity improvement. In addition, we propose to make the resource allocation self-organized using a novel concept of *interfering neighbor set* (Section 3). Our contribution is that with an optimal reuse factor of 3 and only one relay per sector, we do a flexible resource allocation based on localized rules amongst the interfering neighbors, which makes our algorithm self-organized.

We then compare performance of our modified FFR scheme with reuse-1, reuse-3 and soft frequency reuse (SFR) in terms of SINR experienced by users (all users and the edge users), edge user's spectral efficiency and area spectral efficiency of these systems.

Our system model has three sectors with each sector having a critical and non-critical region corresponding to edge and cell-centre users respectively. Resource allocation is performed for critical users using one-third of the resources available in each critical region. Now, the RBs selected for non-critical region (say, region 1D) are those which are orthogonal to the ones allocated in the critical region of that sector (region 1A) and also to the other two non-critical regions (region 1E and 1F) of the same cell. Thus, resource allocation is done such that no channel is given to more than one user belonging to same interfering neighbor set.

The motivation for imposing such restriction on allocation of RBs is to reduce the number of interferers and improve the SINR of all users. This is achieved due to eNB and RN antenna being directional. It has been illustrated in sub-section 4.1 where we discussed the interference scenario for two cases: one without RNs deployed and the other with RN deployed in our system model.

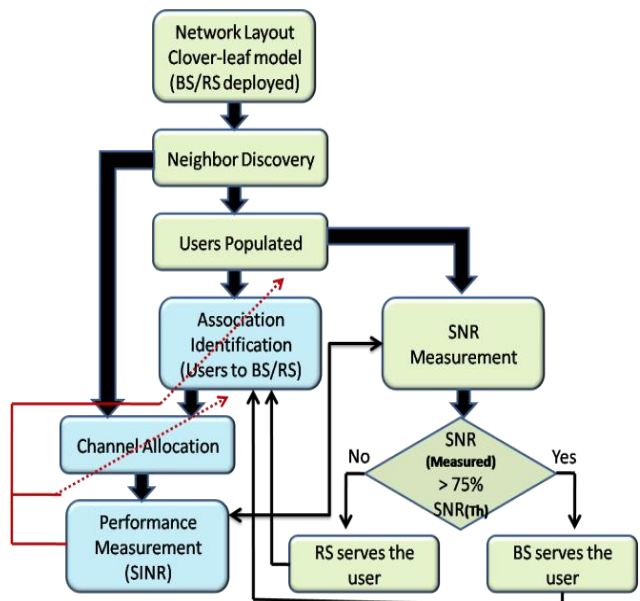

Figure 9: Flowchart of self-organized spectrum allocation in RACN

The flowchart of proposed self-organized resource allocation scheme is shown in Figure 9. Once the network is deployed, we identify the interfering neighbor set for each region as mentioned in Section 4. Then, users are differentiated as cell-center or edge users based on their SNR and accordingly, their serving nodes are identified. Then, based on the interfering neighbor set identification, an orthogonal resource allocation is done within every set of such interfering neighbors (indicatively shown by the colors in Figure 3a). This strategy relies on orthogonal resource allocation in the local neighborhood, which ensures that the adjacent cells are not the co-channel ones. Thus, we avoid the worst-case interference scenario by coordination. This significantly reduces interference and improves system performance.

This self organized scheme is based on the notion of self organization in nature where simple localized rules cascaded over an entire network results in an emergent organized pattern. We thus choose a local set of sectors. Each sector is assumed to have perfect knowledge of its current allocation and user demand as well as that of every sector in its local neighborhood. After implementing the modified FFR scheme, we add another dimension of flexibility by allowing coordination among neighbor sets for resource allocation. This coordination is based on the resources available, interference levels and the user demand.

## 6. Simulation Results and Performance Analysis

The simulations are performed for OFDMA downlink transmission in the framework of 3GPP-LTE. A few assumptions made in this simulation are:

1. Perfect channel state information on the link between eNB and RN is available.
2. Users (also known as User Equipment or UE as per 3GPP-LTE standards) are uniformly distributed.
3. Users have uniform rate requirement.
4. There is no intra-cell interference as OFDMA is used as the radio access technology.
5. There is no inter-sector interference in a cell site.
6. Both eNB and RN employ sectored antennas.

Table1: Simulation Parameters

| Simulation Parameters | |
|---|---|
| System Bandwidth | 10 MHz |
| Sub-channel Bandwidth ($\Delta f$) | 15 kHz |
| Transmit Power eNB ($P^{eNB}$) | 43 dBm |
| Transmit Power RN ($P^{RN}$) | 40 dBm |
| Noise Spectral Density ($N_o$) | -174 dBm/Hz |
| Log-normal shadowing std. deviation eNB-UE ($\sigma_{eNB}$) | 8 dB |
| Log-normal shadowing std. deviation RN-UE ($\sigma_{RN}$) | 6 dB |
| Inter-site distance | 1.5 Km |

Instead of wrap-around model, we consider performance of a reference cell which is the central cell in a seven cell system. It eliminates any edge effects. Simulations are done in MATLAB and simulation parameters are mentioned in the Table 1. We consider log-normal shadowing $\xi$ on each link, where $\xi$ is a Gaussian random variable with mean 0 and standard deviation $\sigma_{eNB}$ and $\sigma_{RN}$ for eNB-UE and RN-UE links respectively. We perform simulations for varying number of users in the range of 50 to 5000 users per sector.

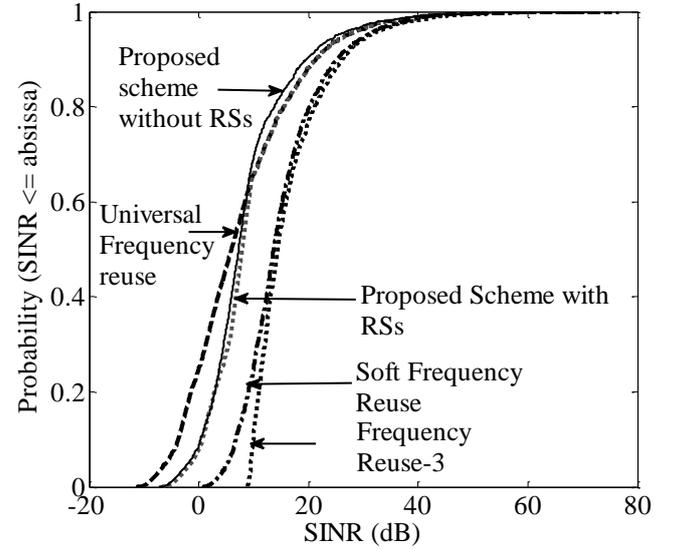

Figure 10: Comparison of the SINR CDF of all users: reuse-1, reuse-3, Soft Frequency Reuse (SFR) and the proposed scheme without and with relays

SINR is measured for all UEs and in particular the cell-edge UEs and its distribution is plotted for reuse-1, reuse-3, SFR, proposed resource allocation scheme without and with relays (Figure 10).

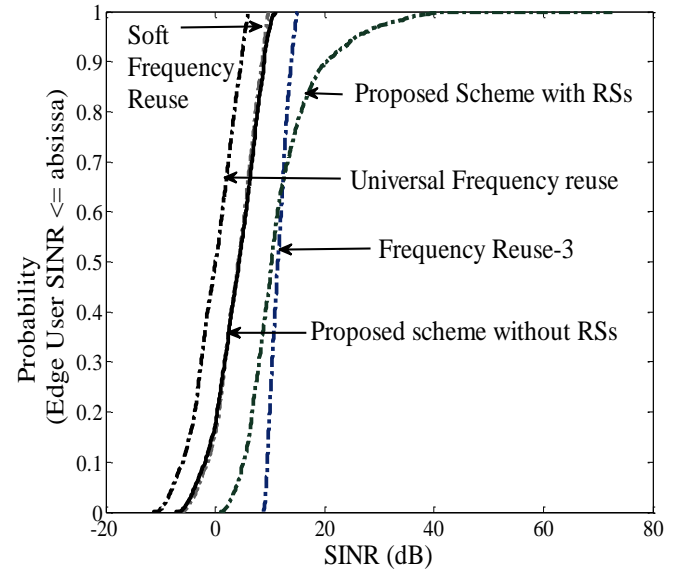

Figure 11: Comparison of the SINR CDF of edge users: reuse-1, reuse-3, Soft Frequency Reuse (SFR) with our proposed scheme without and with relays

It is clearly observed that there is an improvement in SINR performance of all users in the proposed scheme compared to reuse-1, reuse-3 and SFR schemes.

The SINR distribution for edge UEs in the proposed scheme performs better than all other schemes (Figure 11). Also, there is reduction in interference in reuse-3 compared to reuse-1 (Figure 10), albeit the resources available in reuse-3 reduce by a factor of 1/3.

From the histogram plot of SINR of cell edge UEs for all reuse schemes in consideration (Figure 12), it is observed that reuse-3 ensures more number of UEs to experience better SINR compared to reuse-1. It further improves in SFR case and the 'proposed scheme without relays' perform equivalently in this regard. However, a significant improvement is observed in the proposed scheme with relays as large number of users experience better and much higher SINR compared to all other schemes.

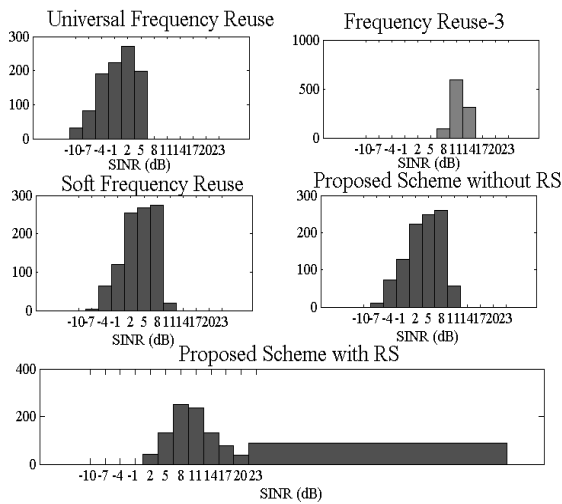

Figure 12: Histogram plot of SINR of the cell edge users for reuse-1, reuse-3, Soft Frequency Reuse (SFR), Proposed scheme without and with relays

The cell edge spectral efficiency is compared for all the schemes (Figure 13) and our proposed scheme outperforms rest other schemes. The area spectral efficiency (Figure 14) for reuse-1 case is the lowest where the entire cell uses all available RBs. It improves in case of reuse-3 where each sector uses a disjoint set of RBs and ensures that edge users encounter less interference compared to reuse-1 case.

The area spectral efficiency improves significantly for SFR case because of the transmit power adaptation and hence, improves the achievable throughput of users. The proposed scheme without relays gives higher area spectral efficiency compared to reuse-1 and reuse-3 because the non-critical region is also sectored into three regions. However, it is slightly lesser than the SFR as there is no power adaptation and the transmit power switches between only two fixed power levels. Our proposed resource allocation scheme with RNs outperforms all other schemes.

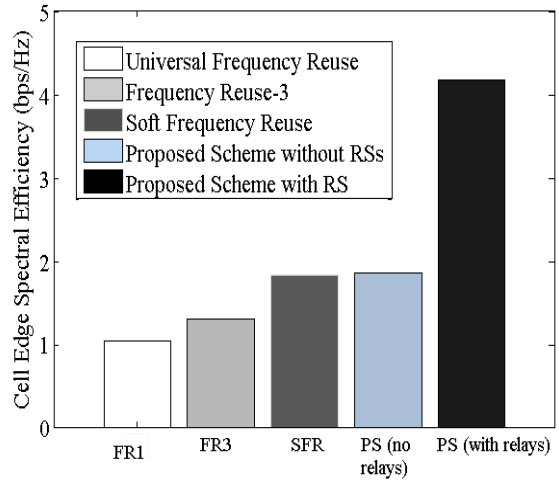

Figure 13: Comparison of the Cell Edge Spectral Efficiency for: reuse-1, reuse-3, Soft Frequency Reuse (SFR) and the proposed scheme without and with relays

However, there exist a few limitations of the proposed scheme as increased overheads due to information exchange between entities will consequently increase computational complexity at RN. Also, it does not allow exploiting multi-user diversity as discussed in Section 2.2.

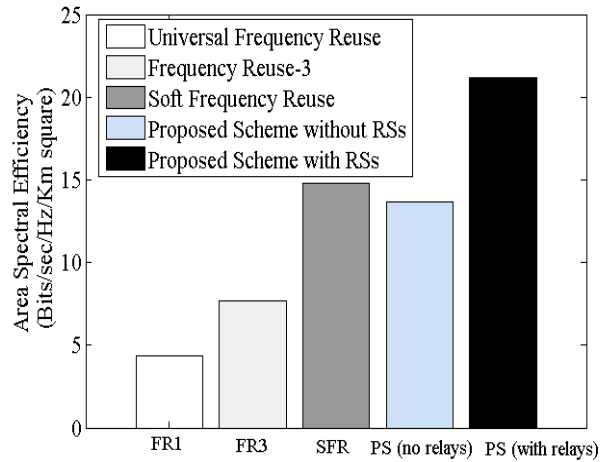

Figure 14: Comparison of the Area Spectral Efficiency of the system: reuse-1, reuse-3, Soft Frequency Reuse (SFR) with our proposed scheme (without and with relays)

## 7. Conclusions and Future Work

In this paper, we reviewed the resource planning schemes in OFDMA-based cellular networks and discussed the significance of channel partitioning schemes like FFR, SFR over the traditional reuse plans. We also investigated the work done for ICI mitigation in relay-assisted cellular networks via dynamic and self-organized approaches available in the literature. We went further to introduce our proposed self-organized resource allocation scheme with ICIC and showed from simulation results that our scheme performs better for the edge users in the DL transmission of an OFDMA-based RACN. We introduced a novel concept of *interfering neighbor set* in which resource allocation decision is taken by coordinating with entities locally. It helps in achieving improved system spectral efficiency and edge users' performance by reducing ICI. The distributed nature of algorithm (due to localized interaction between entities) makes it simple to implement and the dynamic nature ensures efficient resource utilization. Finally the results exhibits that our proposed self-organized resource allocation scheme with relays outperforms the existing schemes by providing higher SINR values for a large proportion of edge users without affecting the overall system performance.

In our system model, relay placement at the cell edge is done with a foresight that in future, we will make the RNs self-organized by facilitating them to switch their association between the neighboring eNBs based on the traffic load in a sector and the serving capacity of RN. This will improve system efficiency even when there is variable rate requirement of users in a non-uniform traffic distribution scenario and also achieve load balancing.

## 8. Acknowledgement

This project is being carried out under the India-UK Advanced Technology of Centre of Excellence in Next Generation Networks (IU-ATC) project and funded by the Department of Science and Technology (DST), Government of India and UK EPSRC Digital Economy Programme.